\documentclass[letterpaper]{article} 
\usepackage{aaai25}  
\usepackage{times}  
\usepackage{helvet}  
\usepackage{courier}  
\usepackage[hyphens]{url}  
\usepackage{graphicx} 
\usepackage{natbib}  
\usepackage{caption} 
\usepackage{algorithm}
\usepackage{makecell}
\usepackage[noend]{algpseudocode} 

\usepackage{subcaption}  
\usepackage{hhline}
\usepackage{multirow}
\usepackage{placeins}
\usepackage{amssymb}

\usepackage{diagbox}
\usepackage{booktabs}
\usepackage{array}

\usepackage{amsthm}
\usepackage{amsmath}

\usepackage[table,xcdraw]{xcolor}
\usepackage[modulo,switch]{lineno} 

\usepackage{newfloat}
\usepackage{listings}
\DeclareCaptionStyle{ruled}{labelfont=normalfont,labelsep=colon,strut=off} 
\floatstyle{ruled}
\newfloat{listing}{tb}{lst}{}
\floatname{listing}{Listing}
\title{Real-Time LaCAM for Real-Time MAPF}
\author{
    Runzhe Liang\equalcontrib\textsuperscript{\rm 1}, Rishi Veerapaneni\equalcontrib\textsuperscript{\rm 1}, Daniel Harabor\textsuperscript{\rm 2}, Jiaoyang Li\textsuperscript{\rm 1}, Maxim Likhachev\textsuperscript{\rm 1}
}
\affiliations{
    \textsuperscript{\rm 1} Carnegie Mellon University\\
    \textsuperscript{\rm 2} Monash University\\
    \{runzhel, rveerapa, jiaoyanl, mlikhach\}@andrew.cmu.edu, daniel.harabor@monash.edu 
}

\usepackage{bibentry}

\begin{document}

\maketitle

\begin{abstract}
The vast majority of Multi-Agent Path Finding (MAPF) methods with completeness guarantees require planning full-horizon paths. However, planning full-horizon paths can take too long and be impractical in real-world applications. Instead, real-time planning and execution, which only allows the planner a finite amount of time before executing and replanning, is more practical for real-world multi-agent systems. 
Several methods utilize real-time planning schemes but none are provably complete, which leads to livelock or deadlock. Our main contribution is Real-Time LaCAM, the first Real-Time MAPF method with provable completeness guarantees. We do this by leveraging LaCAM in an incremental fashion. Our results show how we can iteratively plan for congested environments with a cutoff time of milliseconds while still maintaining the same success rate as full-horizon LaCAM. We also show how it can be used with a single-step learned MAPF policy. 

\end{abstract}

\section{Introduction}
Multi-Agent Path Finding (MAPF) is the problem of finding collision-free paths for a team of agents in a shared congested environment. MAPF is particularly relevant for warehouse automation, which requires dozens to hundreds of robotic agents to navigate effectively.

The majority of MAPF methods focus on finding a full-horizon solution quickly. However, real-world applications have strict planning time limits. In these scenarios, real-time planning and execution is required. In this setup, the planner has a fixed (small) planning budget to compute the next action for all the agents to take. The agents then take this action and repeat the planning and execution process.

The most prevalent framework for real-time planning and execution is to utilize windowed planning, where instead of computing an entire collision-free path, the planner computes a partial collision-free path for the next $W$ time steps. Agents then move along this path to some extent and replan. Windowed planning decreases the planning time to fit within small realistic planning budgets.

However, the vast majority of real-time and windowed MAPF methods are theoretically incomplete and, in practice, can suffer from deadlock or livelock. Recent work introduced the Windowed Complete MAPF (WinC-MAPF) framework that enables completeness using heuristic updates and agent groups, but can empirically take several seconds to plan a single step action for all agents and is hence not practically real-time \cite{veerapaneni2024winc_mapf}. 

Thus, to our knowledge, there exists no theoretically complete real-time MAPF method. To that end, we design Real-Time LaCAM. Real-Time LaCAM incrementally builds the LaCAM depth-first search \cite{okumura2023lacam} and is complete. We empirically show how Real-Time LaCAM can have an identical success rate and overall runtime of full-horizon LaCAM with a per-iteration cutoff time of milliseconds (or smaller). We additionally show how it can be used with a learned machine learning MAPF policy.

\section{Related Works}
\subsection{Real-Time MAPF Formulation}
Multi-Agent Path Finding (MAPF) involves finding collision-free paths for a set of $N$ agents, denoted as ${i = 1, \dots, N}$, where each agent must travel from its start location $s_i^{\text{start}}$ to its goal location $s_i^{\text{goal}}$. In the standard 2D MAPF setup, the environment is discretized into grid cells and timesteps. Agents can move to adjacent cells in any cardinal direction or remain stationary in their current cell. 

We define the MAPF search problem in the joint configuration space. A joint configuration at timestep $t$, $C^t$, is the location of all agents at timestep $t$, i.e. $C^t = [s^t_1, s^t_2, ..., s^t_N]$. A valid MAPF solution is a sequence of joint configurations $\Pi = C^0, C^1, ..., C^T$ where for all agents $i$, $C^0_i = s_i^{start}$ and $C^T_i = s_i^{goal}$. 
To ensure validity, the solution must avoid vertex collisions (when two agents occupy the same cell at the same timestep i.e., $C^t_i = C^t_j$ for $i \neq j$) and edge collisions (when two agents swap positions between consecutive timesteps, i.e., $C^t_i = C^{t+1}_j \wedge C^{t+1}_i=C^t_j$, for all timesteps $t$). The standard objective in MAPF is to find a solution $\Pi$ that minimizes the total cost $|\Pi| = \sum_{i=1}^N \sum_{t=0}^{T-1} c(C_i^t,C_i^{t+1})$. In this work, we assume that all actions are of unit cost, $c(C_i^t,C_i^{t+1})=1$, except when an agent remains at its goal (where the cost is 0). 

Although the above describes the standard formulation of the full-horizon MAPF problem, this work focuses on real-time planning where the planner does not have unlimited time to find a solution. Instead, agents iterate through planning and execution. 
At every planning iteration, the objective is to determine the next configuration $C^1$ to move to (where $C^0$ is the current configuration of the agents). This is done by employing a time-bounded search that finds the best $\Pi^{0:W} = C^0, C^1, ..., C^W$ collision-free partial path where $W$ is the length of the partial path the search reached when reaching the timeout (and is not a fixed constant). Agents move to $C^1$ and then repeat. 
Real-Time planning is closely related to windowed planning but has an important subtle difference: windowed planning computes a partial path to a predefined fixed $W$ and may not satisfy a fixed timeout. In practice, real-time methods usually adjust their window size so that it empirically runs within their target runtime limit.



\subsection{Real-Time and Windowed MAPF Solvers}
There are a variety of MAPF solvers that utilize windowed planning. Windowed Hierarchical Cooperative A* \cite{cooperativeSilver2005} utilized a modified prioritized planner \cite{erdmann1987multiple} that planned collision-free paths for only $W$ timesteps. Rolling-Horizon Collision Resolution \cite{rhcrLi2020} generalizes this idea by modifying a variety of modern solvers (CBS \cite{sharon2015cbs}, ECBS \cite{barer2014suboptimalecbs}, and PBS \cite{pbs2019}) to plan partial paths.

The introduction of the Robot Runners competition, which requires planning for 1000's of agents in 1 seconds, empirically requires real-time windowed solvers. The 2023 winning solution Windowed Parallel PIBT-LNS (WPPL) \cite{jiang2024scaling_mapf_competition} planned an initial partial path using PIBT \cite{okumara2022pibt_jair} and then refined it with the remaining planning time using parallel LNS \cite{li2021mapf-lns,li2022mapf-lns2}.

All these prior real-time/windowed approaches are theoretically incomplete and mention deadlock/livelock as a problem. Recently, the Windowed Complete MAPF (WinC-MAPF) framework showed how to enable completeness for windowed planners \cite{veerapaneni2024winc_mapf} by leveraging heuristic updates from single-agent real-time heuristic search \cite{korf1990_lrta}. However, they require an optimal windowed solver (which minimizes $|\Pi^{0:W}|$ incorporating the heuristic updates), which in practice means it can take several seconds to find a one-step path.

Our method shows an alternative method for real-time/windowed search with completeness by modifying LaCAM. Our method does not require optimal solvers and thus can use PIBT and plan for near arbitrarily small timeouts.

\begin{figure*}[t]
    \centering
    \includegraphics[width=0.9\linewidth]{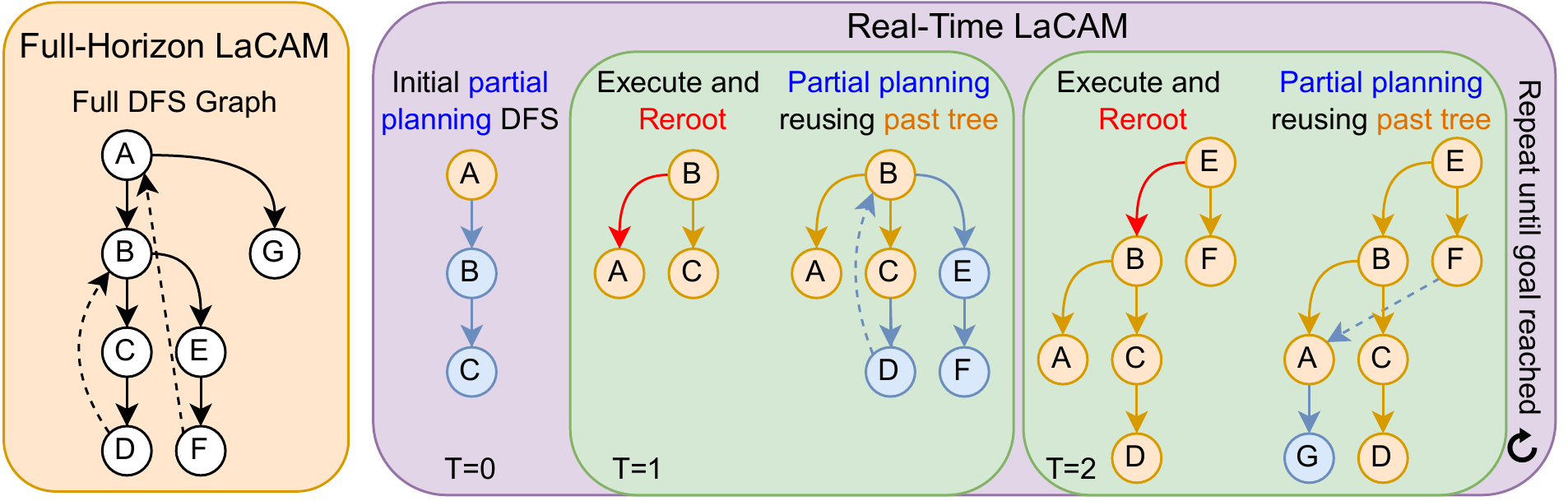}
    \caption{Left: The full-horizon DFS required for LaCAM to find a solution. Letters denote joint configurations while arrows denote transitions (via PIBT and constraints). Right: Instead of needing to plan the entire horizon, Real-Time LaCAM incrementally builds up the DFS. T=0: First, there is an initial partial LaCAM search that terminates once it reaches the timeout. T=1: Then, the agents move to the next configuration ($A \rightarrow B$). Importantly, we re-root the tree by swapping the $A \rightarrow B$ edge so that the DFS tree is now rooted at the new location $B$. We then plan the next iteration reusing the DFS tree, in this case starting at $C$ (the end of initial DFS). If we revisit configurations from previous searches ($B$ in this example), we add constraints. This prevents deadlock/livelock and ensure completeness, compared to naively re-searching from scratch which would enter a loop in this case. T=2+: This process repeats across timesteps until the goal is reached.}
    \label{fig:overview}
    \vspace{-0.5em}
\end{figure*}

\subsection{Real-Time Single-Agent Heuristic Search}
There are a variety of single-agent heuristic search algorithms specifically designed for real-time search and execution. In particular, Learning Real-Time A* (LRTA*) \cite{korf1990_lrta} showed how updating the heuristic of states that the agent visits can ensure completeness. This approach was popular and has been extended for better performance \cite{koenig2009_lss_lrta,rivera2013weighted_real_time_search}.

A different less popular approach for completeness is Time-Bounded A* (TBA*) \cite{bjornsson2009time_bounded_astar} which builds a single A* search rooted at the start location across many iterations. At every planning step, the single A* search continues from before. The agent then moves towards the node with the best f-value. Importantly, this could mean moving forward (if that node is a child of the current location), or backwards (if the node is a child of an ancestor). In the worst case, an agent in TBA* could need to backtrack all the way to the start location, but is still eventually guaranteed to reach the goal as the single A* search is complete. Our Real-Time LaCAM method can be interpreted as a MAPF version of TBA* using LaCAM's search tree instead of an A* search.

\subsection{LaCAM}
LaCAM \cite{okumura2023lacam} is an extremely fast MAPF solver that utilizes a \textit{lazy} Depth-First Search (DFS) over configurations. We note that a regular DFS is infeasible as it requires generating on the order of $5^N$ possible successor configurations when expanding a configuration.

\textbf{Lazy DFS:} LaCAM starts with a configuration $C^0$, and instead of generating all valid successor configurations, it only generates one successor $C^1$. It repeats this process of generating only one successor for each configuration. Unlike a regular DFS, LaCAM allows revisiting previously generated configurations. Crucially, if a $C^k$ is revisited from $C^{k'}$, LaCAM adds a constraint to the success generation and requires $C^k$ to generate a new successor (i.e., different from $C^{k+1}$). Thus, LaCAM's DFS is a tree with \textit{backjumps} that enables the DFS to branch on previously visited configurations rather than a typical DFS which only revisits configurations when backtracking. 

LaCAM imposes constraints lazily as well by iteratively constraining agents' actions. The first 5 times that a configuration $C^k$ is revisited, the first agent is constrained to each of its 5 different actions. The next $5^2$ times, the first two agents are constrained to the different combinations of their 5 actions. This logic repeats where, in the worst case, all $5^N$ different neighboring configurations are explicitly generated.

\textbf{Configuration Generator:} Although the ``configuration generator" in LaCAM can be any one-step MAPF method, in practice it needs to be very fast, and thus Priority Inheritance with Backtracking (PIBT) is used \cite{okumara2022pibt_jair}. For this work, we do not need to know PIBT's internals but only that it is a very fast one-step MAPF planner.


\textbf{Completeness:} LaCAM has completeness guarantees if it is finding a full-horizon solution as it eventually searches the entire state-space. However, if LaCAM is iteratively run in a time-bounded or windowed fashion where it does not find a solution to the goal, it is not complete as the DFS's between different iterations could search the same state-space and get stuck in deadlock/livelock.

\section{Real-Time LaCAM}
Our main insight is that instead of running a single full-horizon LaCAM DFS, we can incrementally build up (and execute) the LaCAM DFS through repeated calls that remember past history.
Conceptually, we can imagine maintaining a ``global" DFS tree that we build up across iterations. At every iteration of planning, we continue the DFS from where we left off. When we reach the per-iteration planning cutoff, we backtrack from the latest configuration in the DFS to the current configuration to find the current path and the next configuration to move to.

The main problem with this description is that since we are moving along the DFS tree across iterations, it is possible that when backtracking to reconstruct the path, we will \textit{not} encounter the current configuration.
Our solution is simple: reroot the global tree so that the current configuration is always the root. This ensures that backtracking from any configuration will always reach the current configuration. When moving from a configuration $C^A \rightarrow C^B$, we just swap the parent pointer so that $C^A$'s parent is now $C^B$. Note that this relies on the fact that MAPF graphs are bidirectional, which is true in current applications.

\textbf{Example:}
Figure \ref{fig:overview} depicts the Real-Time process. We start at configuration $A$ and initially plan up to $C$ until our per-iteration timeout is reached. The next configuration according to this partial plan is $B$, so the agents move to $B$ at the next timestep.

\textbf{T=1:} Since we moved to $B$, we need to reroot the current DFS tree to $B$ by reversing the executed $A \rightarrow B$ edge (highlighted in red). We then proceed to plan by continuing the DFS from where it left off, in this case starting at $C$. By maintaining the global DFS, the per-iteration planning can remember revisiting a configuration ($B$ in this case) and generate a new successor ($E$). This contrasts with running LaCAM from scratch at each iteration, where it could get stuck revisiting the same configurations across different iterations. When planning reaches its timeout when reaching $F$, we backtrack $F, E, B$ and accordingly move to $E$.

\textbf{T=2:} We repeat the process, rerooting the tree at $E$ and continuing the DFS from $F$. Again, if the DFS revisits a configuration, it adds constraints and generates a new configuration.

\subsection{Theoretical Properties} \label{sec:theoretical-properties}
The main observation is that Real-Time LaCAM builds up an identical search tree as LaCAM (across iterations instead of all at once) except for rerooting. However, rerooting does not change the search configurations or constraints.

Thus, Real-Time LaCAM is complete as full-horizon LaCAM is complete (as it will eventually search the entire configuration space). Additionally, Real-Time LaCAM has a near identical overall planning time as the full-horizon planning time (i.e., the sum of planning time across all planning iterations will equal full-horizon LaCAM's planning time) as it builds the same tree. 
The only differences are the rerooting and backtracking operations, which have negligible runtimes compared to other operations.

\begin{figure*}[t]
    \centering
    \includegraphics[width=0.95\linewidth]{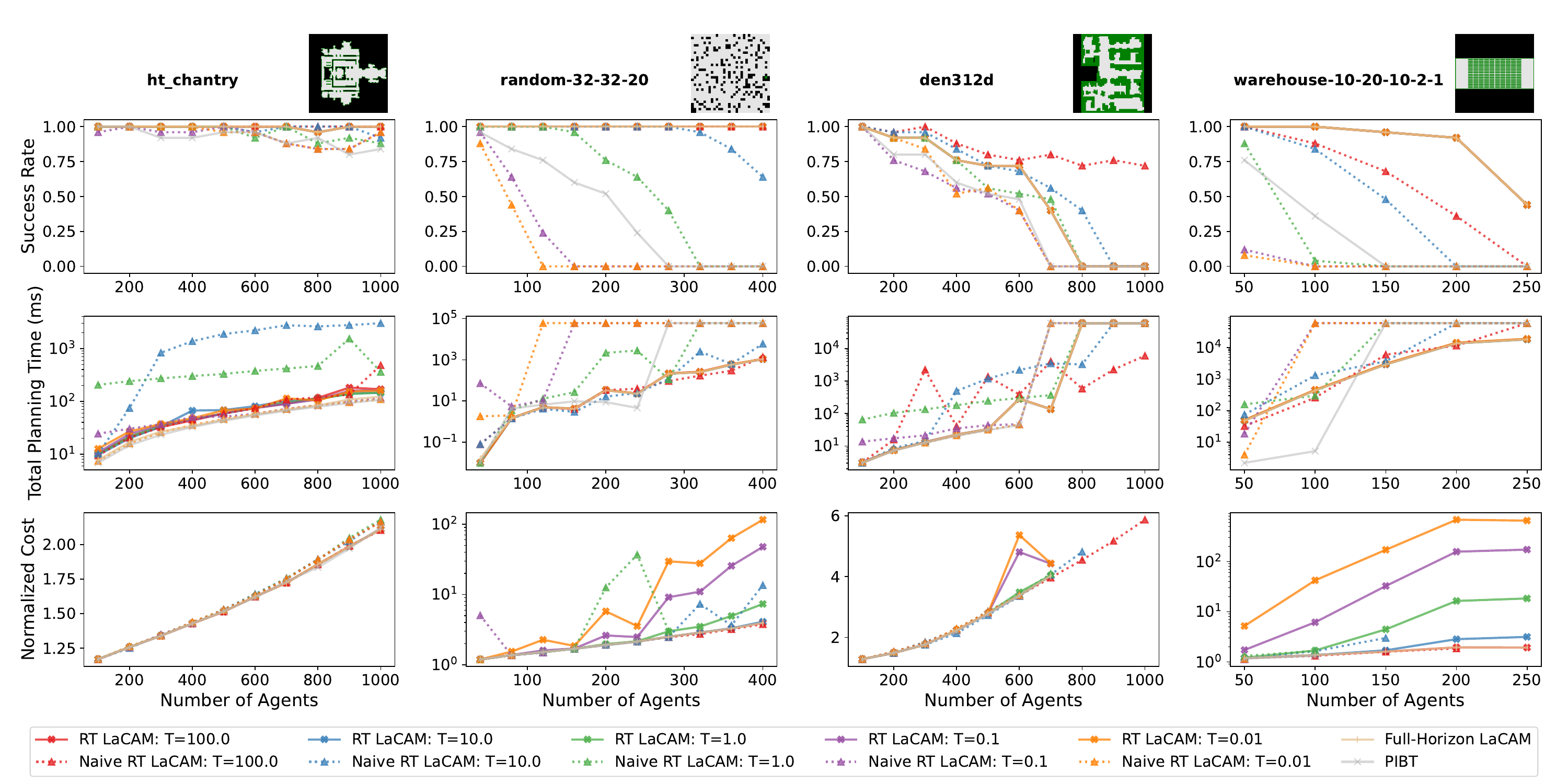}
    \caption{Comparing Real-Time LaCAM (RT LaCAM) with Naive Real-Time LaCAM (Naive RT LaCAM) with different per-iteration cutoff times in milliseconds. All Real-Time LaCAM have identical success rates (middle row) and total planning time (top) as full-horizon LaCAM and thus perfectly overlap in those plots.
    Real-Time LaCAM, especially with small timeouts (orange, purple, green), has better success rates than Naive Real-Time LaCAM.}
    \label{fig:main-results}
    \vspace{-0.5em}
\end{figure*}

\section{Experimental Results}
We compare Real-Time LaCAM with naive Real-Time LaCAM (which replans from scratch at every iteration), PIBT, and full-horizon LaCAM on a subset of the standard benchmark \cite{stern2019mapfbenchmark}, with 25 scenes per map in Fig \ref{fig:main-results}. Real-Time LaCAM (ours and naive) are run with per-iteration cutoffs of $T=0.01, 0.1,1,10,100$ milliseconds. These methods and PIBT interleave planning and execution, while full-horizon LaCAM only plans once. Real-Time LaCAM methods execute the first action of their planned partial path at every iteration. Methods are run with a cumulative 60-second planning timeout. The second row is the sum of the planning time across all iterations of iterative planning and action execution (with timed-out runs included as 60 seconds). The third row plots the normalized solution cost (raw solution / sum of agent's individual optimal path).

We first see, or more precisely, struggle to visually see, Real-Time LaCAM in the success rate or runtime plots (first two rows). Upon close inspection, we see that all Real-Time LaCAM methods (independent of the timeout) perfectly overlap with full-horizon LaCAM. This verifies our theoretical properties (Sec \ref{sec:theoretical-properties}) that Real-Time LaCAM builds an identical DFS tree to full-horizon LaCAM. The main effect of per-iteration cutoffs is on the solution quality. In particular, Real-Time LaCAM with small cutoffs (e.g. 0.01, 0.1 ms) in random-32-32-20 and warehouse has 10-100x worse solution quality due to their myopic planning.

Naive Real-Time LaCAM has more varying results.  First, we see that with small cutoffs (0.01, 0.1, 1 ms) its success rate on tough maps (random-32-32-20, warehouse) is extremely poor. Second, the cumulative planning time varies substantially between different iteration cut-offs.



\begin{figure}[t!]
    \centering
    \includegraphics[width=0.65\linewidth]{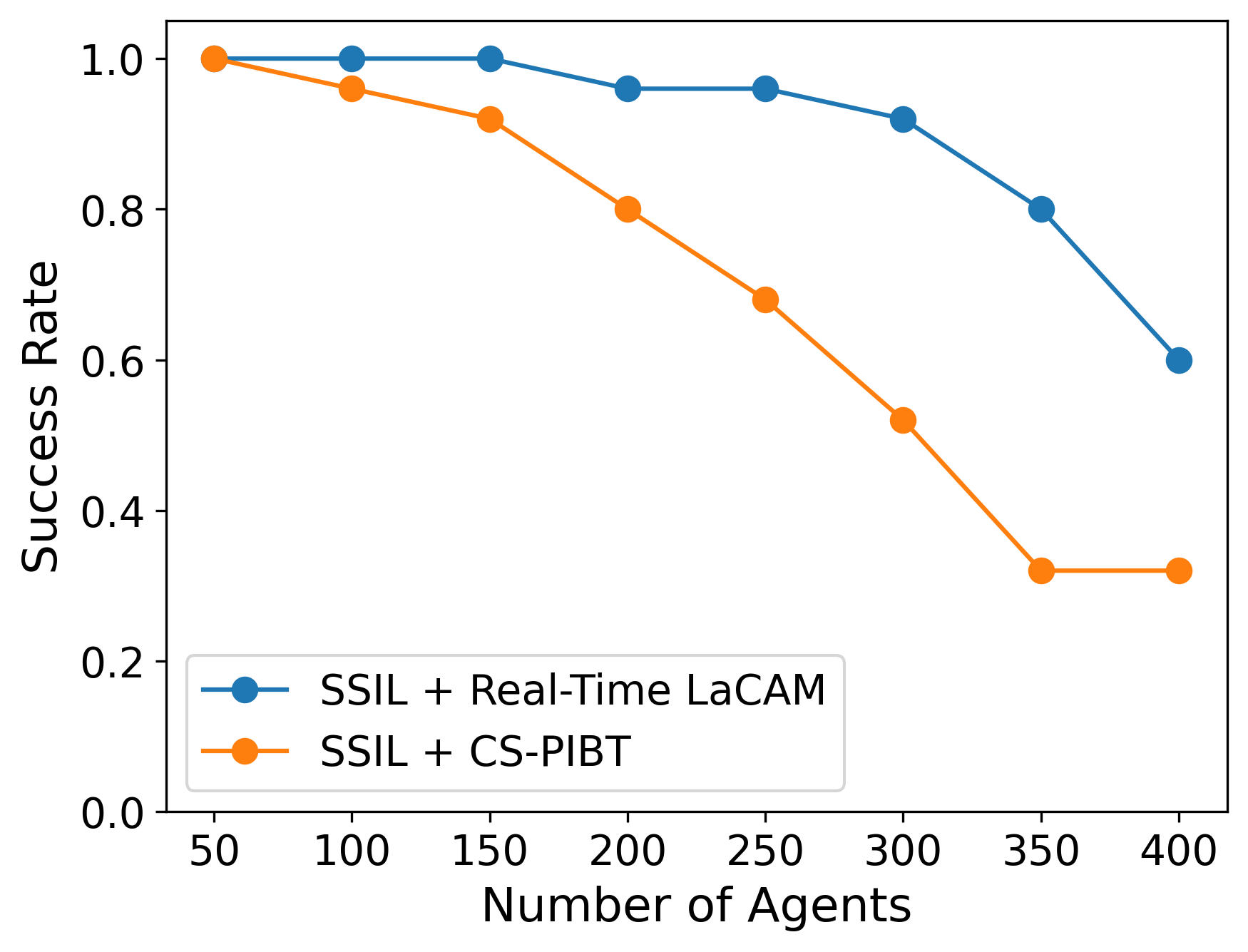}
    \vspace{-0.5em}
    \caption{Running a pretrained single-step ML MAPF policy on random-32-32-20 with Real-Time LaCAM vs CS-PIBT.}
    \label{fig:nn-results}
    \vspace{-1em}
\end{figure}

Additionally, Real-Time LaCAM can directly be used with learned MAPF policies. These policies predict a next action probability distribution per agent which can be post-processed using collision-shield PIBT (CS-PIBT) \cite{veerapaneni2024improving_mapf_policies_with_search}. Conceptually, the model's predictions replace the backward Dijkstra heuristic of PIBT. Thus, we can directly use Real-Time LaCAM with a policy and CS-PIBT. Fig \ref{fig:nn-results} shows how post-processing a pretrained model SSIL \cite{veerapaneni2024work_smart_not_harder} with Real-Time LaCAM improves performance compared to CS-PIBT.

\section{Future Work and Conclusion}
Identical to regular LaCAM, Real-Time LaCAM works for any configuration generator that can incorporate constraints. We note that incorporating constraints in MAPF is trivial as it just forces the $k$ constrained agents to move to specific positions and reduces the configuration generation problem from $N$ agents to the $N-k$ unconstrained agents. Thus, Real-Time LaCAM can be viewed as a framework for taking any MAPF planner and making it complete in windowed planning.
This broader perspective offers a different avenue for windowed completed MAPF planners compared to the WinC-MAPF framework. In particular, the WinC-MAPF framework requires an optimal solver, heuristic updates, and computing of disjoint agent groups. We solely require applying constraints. Conceptually, a heuristic update says that a configuration is expensive but does not specify which agents should move or that the search should avoid that configuration. On the flip side, constraints explicitly dictate which agents move and can explore new configurations faster.

One promising future work is to try to merge the ideas of using constraints from LaCAM and heuristic updates from WinC-MAPF. This could enable fast solvers while maintaining better solution qualities.
Additionally extending Real-Time LaCAM to Engineering LaCAM* \cite{okumara2024engineering_lacam} could produce higher quality real-time results.


Overall, we introduce Real-Time LaCAM, the first real-time MAPF method with completeness guarantees. We show how it has an impressive success rate with tiny (milliseconds) per-iteration cutoffs compared to existing methods that get stuck in deadlock/livelock, and can be used with a learnt policy. 

\section*{Acknowledgments}
This work was partially supported by National Science Foundation (NSF) grant \#2328671, NSF Graduate Research Fellowship Program grant \#DGE2140739, and a gift from Amazon.

\bibliography{main_socs}

\end{document}